\pdfoutput=1

\documentclass[11pt,twoside,a4paper,cmspaper,final,collab]{cms-tdr}
\def\svnVersion{166417}\def\cmsCernNo{2012-100}\def\cmsCernDate{\today}\def\cmsMessage{Submitted to the Journal of High Energy Physics}

\begin{document}\cmsNoteHeader{EXO-11-036}

\hyphenation{had-ron-i-za-tion}
\hyphenation{cal-or-i-me-ter}
\hyphenation{de-vices}

\RCS$Revision: 114342 $
\RCS$HeadURL: svn+ssh://svn.cern.ch/reps/tdr2/papers/EXO-11-036/trunk/EXO-11-036.tex $
\RCS$Id: EXO-11-036.tex 114342 2012-04-04 22:18:24Z alverson $
\newcommand{\cPqbp}{\ensuremath{\cmsSymbolFace{b}^\prime}\xspace} 
\newcommand{\cPaqbp}{\ensuremath{\overline{\cmsSymbolFace{b}}{}^\prime}\xspace} 
\newcommand{\bpbpbar}{\cPqbp\cPaqbp\xspace}

\cmsNoteHeader{EXO-11-036} 
\title{Search for heavy bottom-like quarks in 4.9\fbinv of pp collisions at $\sqrt{s} = 7$\TeV}

\date{\today}

\abstract{
    Results are presented from
    a search for heavy bottom-like quarks, pair-produced in \Pp\Pp\ collisions at $\sqrt{s} = 7$\TeV,
    undertaken with the CMS experiment at the LHC.
    The \cPqbp quarks are assumed to decay exclusively to \cPqt\PW.
    The $\bpbpbar \to \cPqt\PWm\cPaqt\PWp$
    	process can be identified by its distinctive signatures of
    three leptons or two leptons of same charge, and at least one b-quark jet.
    Using a data sample corresponding to an integrated luminosity of 4.9\fbinv,
    observed events are compared to the standard model background predictions, and
    the existence of \cPqbp quarks having masses below
    611\GeVcc is excluded at 95\% confidence level.
}

\hypersetup{%
pdfauthor={CMS Collaboration},%
pdftitle={Search for heavy bottom-like quarks in 4.9 inverse femtobarns of pp collisions at sqrt(s) = 7 TeV},%
pdfsubject={CMS},%
pdfkeywords={CMS, physics, exotica}}

\maketitle 

\section{Introduction}

The total number of fermion generations is assumed to be three in the standard model (SM),
though the model does not provide an explanation of why this should be the case.
Thus the possible existence of a fourth generation remains an important subject for experimental investigation.
Adding a fourth generation of massive fermions to the model
may strongly affect the Higgs and flavour sectors~\cite{Holdom:2009rf, Soni:2010xh,
Buras:2010pi, Erler:2010sk, Flacco:2010rg}.
A fourth generation of heavy quarks would
enhance the production of Higgs bosons~\cite{Denner:2011vt},
while the indirect bound from electroweak precision data on
the Higgs mass would be relaxed~\cite{Frampton:1999xi,Kribs:2007nz}.
Additional massive quarks may provide a key to understanding the matter-antimatter
asymmetry in the universe~\cite{Hou:2008xd}.

Various searches for fourth-generation fermions have already been reported.  Experiments have shown that the number of light neutrino flavours is equal to three~\cite{Decamp:1989tu,Aarnio:1989tv,Adeva:1989mn,Akrawy:1989pi}, but the possibility of additional heavier neutrinos has not been excluded.
A search for pair-produced bottom-like quarks (\cPqbp) by the ATLAS collaboration
excludes a \cPqbp-quark mass of less than 480\GeVcc~\cite{Aad:2012us}.
Earlier studies setting mass limits on possible fourth-generation quarks, from experiments at the Tevatron
and the Large Hadron Collider (LHC), can be found in Ref.~\cite{Aaltonen:2009nr,Aaltonen:2011vr,Chatrchyan:2011em,Abazov:2011vy,Aad:2012bt,Aad:2012xc,CMS_tprime_dilepton}.

Using the Compact Muon Solenoid (CMS) detector, we have searched for
a heavy \cPqbp quark that is pair-produced in
\Pp\Pp\ collisions at a centre-of-mass energy of 7\TeV at the LHC.
We assume that the mass of the \cPqbp quark ($M_{\cPqbp}$) is larger than
the sum of the top quark and the W-boson masses.
If the \cPqbp quark couples principally to the top quark, the decay chain
$\bpbpbar\to \cPqt\PWm\cPaqt\PWp \to \cPqb \PWp\PWm\cPaqb\PWm\PWp$
will dominate~\cite{Arhrib:2006pm}.
Given the 11\% branching fraction for a W-boson to each lepton,
distinctive signatures of $\bpbpbar$ production are expected,
specifically those of two isolated leptons with the same charge (``same-charge dileptons") or
three isolated leptons (``trileptons").  Although occurring very rarely in the standard model,
these two signatures may be present in 7.3\% of the $\bpbpbar$ events.
An earlier search by CMS~\cite{Chatrchyan:2011em} in the same-charge dilepton and the trilepton channels,
utilizing a data set corresponding to an integrated luminosity of 34\pbinv,
set a lower limit on the mass of the \cPqbp quark of 361\GeVcc at the 95\% confidence level (CL).
Here we present an update of this search using a much larger data set, corresponding to an integrated luminosity of 4.9\fbinv.

\section{CMS detector and trigger}

This analysis is based on the data recorded by the CMS experiment in 2011.
The central feature of the CMS detector is a superconducting solenoid,
13 m in length and 6 m in diameter, which provides an axial magnetic field of 3.8\unit{T}.
Charged-particle trajectories are determined using silicon pixel and silicon strip tracker measurements.
A crystal electromagnetic calorimeter,
including lead-silicon preshower detectors in the forward directions,
together with a surrounding brass/scintillator hadronic calorimeter,
encloses the tracking volume and provides energy measurements of electrons and hadronic jets.
Muons are identified and measured in the tracker and in gas-ionization detectors
embedded in the steel return yoke outside the solenoid.
The detector is nearly hermetic, providing measurements of any imbalance of momentum in the
plane transverse to the beam direction.
A more detailed description of the CMS detector can be found in Ref.~\cite{:2008zzk}.

A two-level trigger system~\cite{Adam:2005zf}
selects events for further analysis.
The events analyzed in this search are collected with the requirement
that the trigger system detects at least two lepton candidates.
Efficiencies for these dilepton triggers are determined using events that pass a jet trigger,
have two reconstructed electrons or muons,
and that also pass the full selection criteria described in the next section.
For these selected events, the dilepton trigger efficiencies are estimated to be
91\%, 96\%, and $>$99\%, for events with two muons,
one electron and one muon, and two electrons, respectively.

\section{Selection criteria}

The use of the CMS particle-flow global event reconstruction procedure~\cite{PFT-09-001, PFT-10-001, PFJETPAS, PFT-10-003} has been extended beyond its application in Ref.~\cite{Chatrchyan:2011em}.  In the present analysis, all physics objects -- leptons, jets, and missing transverse energy - are reconstructed with this procedure.
The reconstruction and selection criteria for each physics object used in this analysis are described below.

Candidate muons are reconstructed through a global fit to trajectories,
using hit signals in the inner tracker and in the muon system.
Muons are required to
have transverse momenta $\pt > 20$\GeVc
and $\abs{\eta}<2.4$,
where the pseudorapidity $\eta = -\ln[\tan\theta/2]$ and $\theta$ is the
polar angle relative to the anticlockwise beam direction.
The muon candidate must be associated
with hits in the silicon pixel and strip detectors,
have segments in the muon chambers,
and provide a high-quality global fit to the track segments.
The efficiency for these muon selection criteria is $>$99\% from Z decays~\cite{MUOPAS}.
In addition, the muon track is required to be consistent with
originating from the principal primary interaction
vertex, which is defined by the one associated with tracks yielding the largest value for the sum of their $p_\mathrm{T}^2$.

Reconstruction of electron candidates
starts from clusters of energy deposits in the ECAL, which are
then matched to hits in the silicon tracker.
Electron candidates are required to have $\pt > 20$ \GeVc.
Candidates are required to be reconstructed in the fiducial volume of
the barrel ($|\eta| < 1.44$) or
in the end-caps ($1.57 < |\eta| < 2.4$).
The electron candidate
track is required to be consistent
with originating from the principal primary interaction vertex.
Electrons are identified using variables which include
the ratio between the energies deposited in the HCAL and the ECAL,
the shower width in $\eta$, and
the distance between the calorimeter shower and
the particle trajectory in the tracker,
measured in both $\eta$ and azimuthal angle ($\phi$).
The selection criteria are optimized~\cite{EGMPAS} to
reject the background from hadronic jets while maintaining an
efficiency of 80\% for the electrons from $\rm W$ or $\rm Z$ decays.

Jets are reconstructed by an anti-$k_\mathrm{T}$ jet-clustering algorithm
with a distance parameter $R=0.5$~\cite{Cacciari:2008gp}.
Particle energies are calibrated~\cite{JES} separately for each particle type,
and resulting jet energies therefore require only small corrections that account
for thresholds and residual inefficiencies.
All jet candidates must have $\pt>$25\GeVc and be within $|\eta|<2.4$.
Neutrinos from $\PW$ boson decays escape the detector,
and thereby give rise to a significant imbalance in the net transverse momentum measured for each event.  This missing transverse momentum, expressed as the quantity \ETslash, is defined as the absolute value of the vector sum of the transverse momenta of all reconstructed particles~\cite{MET}.

In contrast to the earlier analysis of Ref.~\cite{Chatrchyan:2011em}, b-tagging is now used
to reject events from backgrounds that do not include a top-quark decay.
The b-tagging algorithm applied in this analysis
generates a list of tracks associated with each jet,
and calculates the significance of each track's impact-parameter (IP),
as determined by the ratio of the IP to its uncertainty.
For the jet to be tagged as a b-jet, the IP significance
of at least three of its listed tracks must exceed a threshold value,
chosen to give an identification efficiency of ~50\%
for b-jets and a misidentification rate of ~1\% for other particle jets~\cite{btagging-performance}.

Electrons and muons from $\PW\to\ell\nu$ ($\ell=\Pe,\mu$) decays
are expected to be isolated from other particles in the detector.
A cone of $\Delta R < 0.3$, where $\Delta R \equiv \sqrt{(\Delta\eta)^2 + (\Delta\phi)^2}$,
is constructed around each lepton-candidate's direction,
and if the scalar sum of the transverse momenta of the particles inside the cone,
excluding contributions from the lepton candidate,
exceeds 15\% of the candidate $\pt$, then the lepton candidate is rejected.
Electron candidates are required to be separated from any selected
muon candidates by $\Delta R > 0.1$
to remove misidentified electrons due to muon bremsstrahlung.
Electron candidates identified as originating from photon conversions are also rejected.

Events are required to have at least one well-reconstructed interaction
vertex~\cite{2010EPJC..tmp..299K}. Events with two leptons of the same electric charge,
or with three leptons (two of which must be oppositely charged), are selected.
For the same-charge dilepton (trilepton) channel,
events with fewer than four (two) jets are rejected.
At least one jet must be identified as a b-jet.
In addition, events that have any two muons or electrons
whose invariant mass $M_{\ell\ell}$ is within 10\GeVcc
of the Z-mass ($|M_{\ell\ell}-M_\cPZ| < 10$\GeVcc) are rejected,
in order to suppress the background from $\cPZ \to \ell^+\ell^-$ decays.
For each event, the scalar quantity
$S_\mathrm{T} = \sum |\vec{p_\mathrm{T}} (\text{jets})| +
\sum |\vec{p_\mathrm{T}} (\text{leptons})| + \ETslash$
is required to  satisfy the condition $S_\mathrm{T} > 500$\GeV.
The selection criteria described above are not fully optimized in terms of
discovery reach, but in fact they are more robust
because they have a single background component in the background estimation with data.

Signal selection efficiencies are estimated using simulated event samples.
Fourth generation quarks production is implemented as a straightforward extension
to the standard model configuration of the \textsc{MadGraph/MadEvent} generator version 5.131~\cite{Maltoni:2002qb}.
Parton showering and hadronization are provided by \PYTHIA 6.424~\cite{PYTHIA} using the matching prescription described in Ref.~\cite{Hoche:2006ph}.
Finally, these generated signal events are passed through the CMS detector simulation based on \GEANTfour~\cite{Agostinelli:2002hh}.

\begin{figure}[thp]
  \begin{center}
    \includegraphics[width=0.45\textwidth]{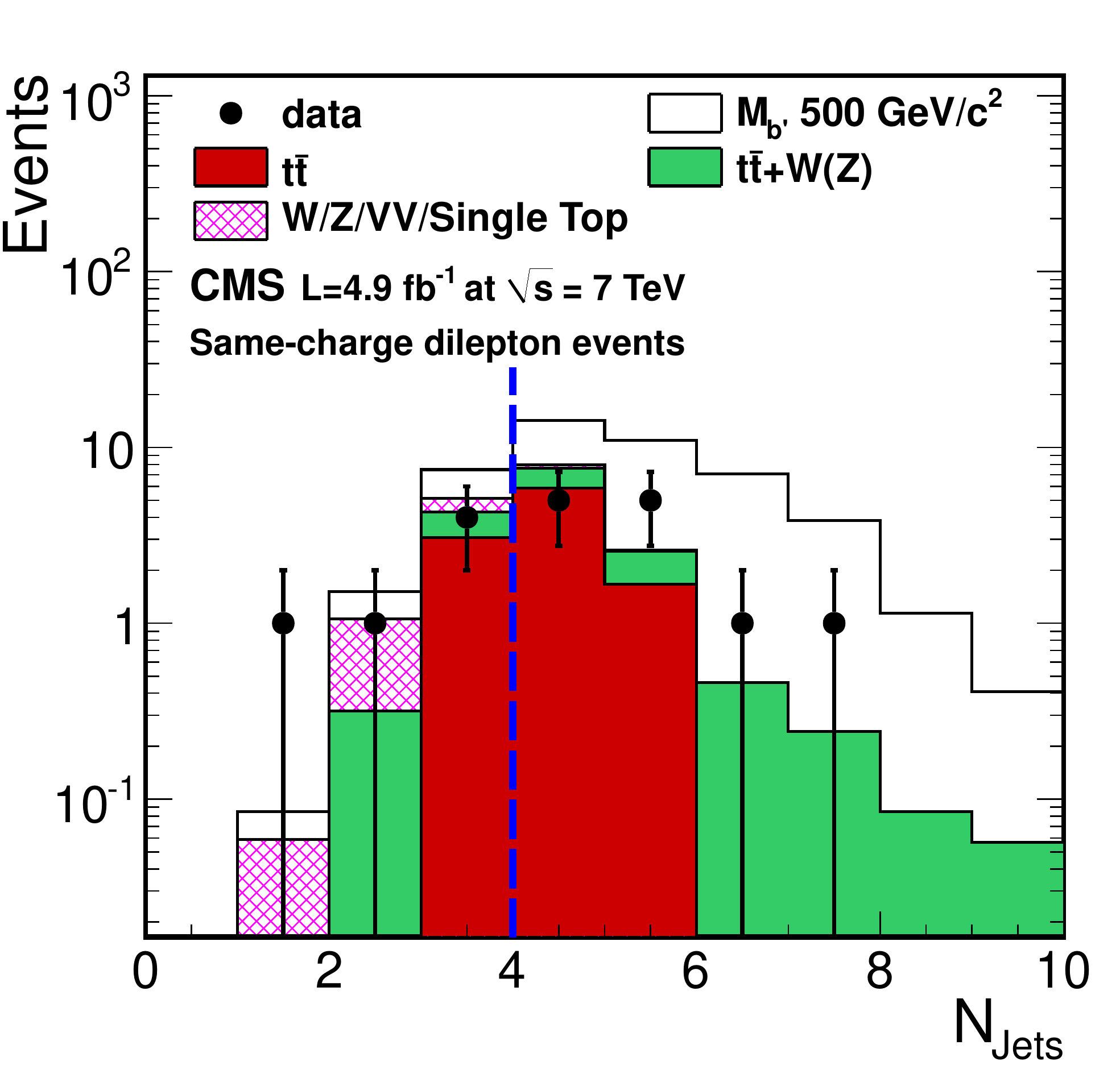}
    \includegraphics[width=0.45\textwidth]{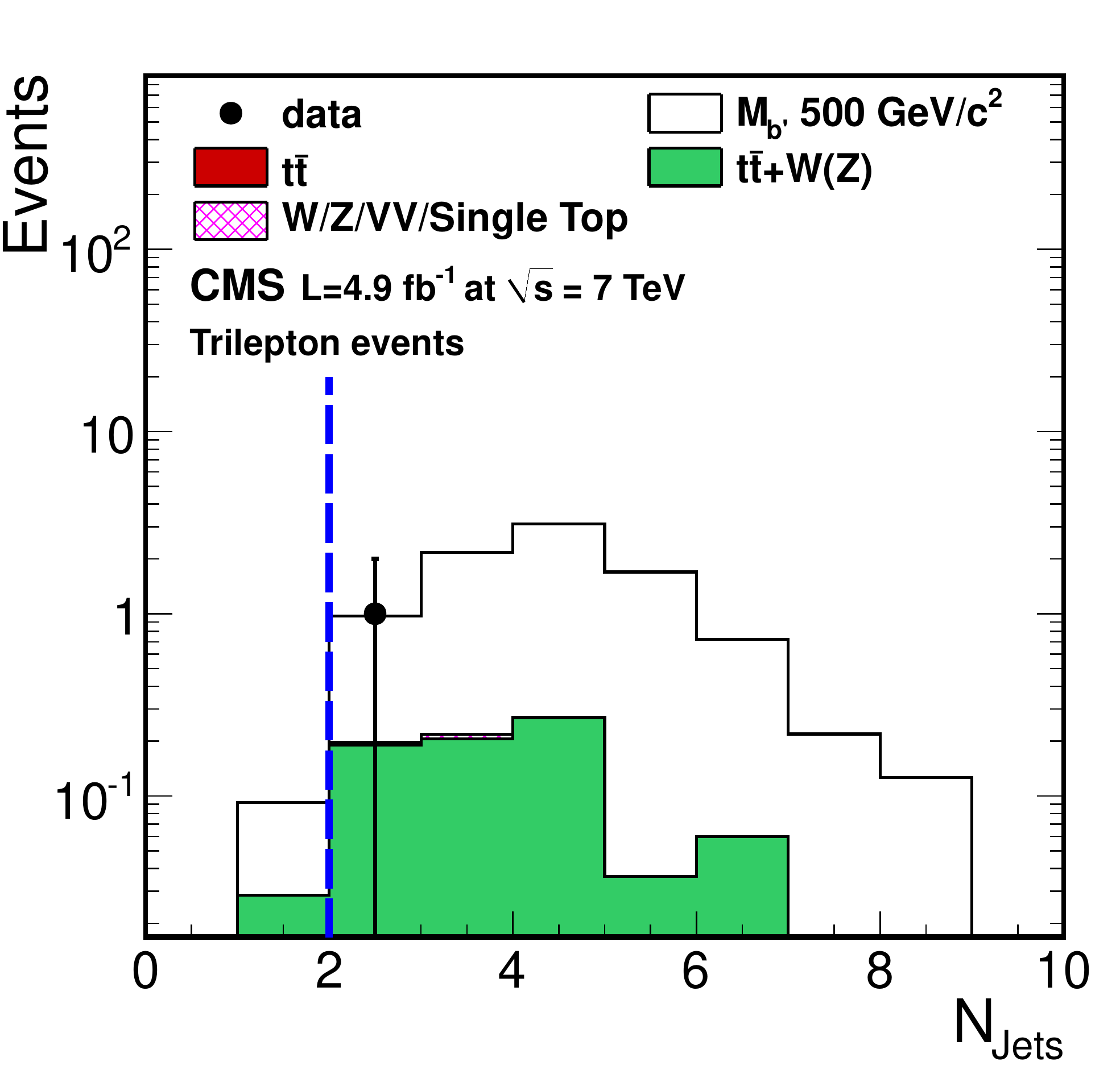}
    \caption{
Jet multiplicity distributions for the same-charge dilepton channel (left), and the trilepton channel (right).
The open histogram shows the contribution expected from a \cPqbp having $M_{\rm b^\prime} =$ 500\GeVcc.
The contributions from standard model processes are normalized to the total estimated background.
All selection criteria are applied except the one corresponding to the plotted variable.
The vertical dotted lines indicate the minimum number of jets required in events selected for each of the channels.
    }
    \label{fig:jet_multiplicities}
  \end{center}
\end{figure}

\begin{figure}[thp]
  \begin{center}
    \includegraphics[width=0.45\textwidth]{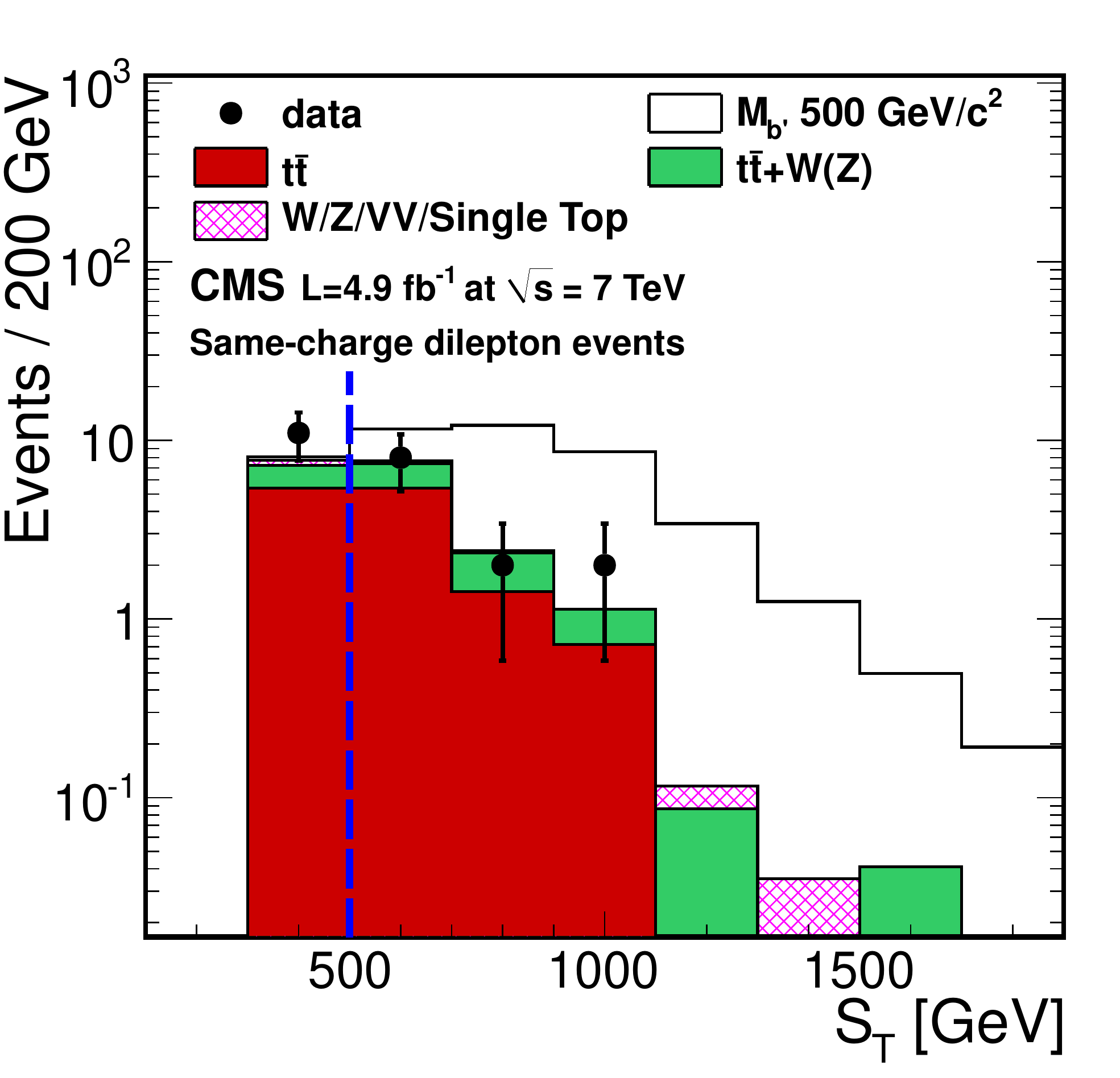}
    \includegraphics[width=0.45\textwidth]{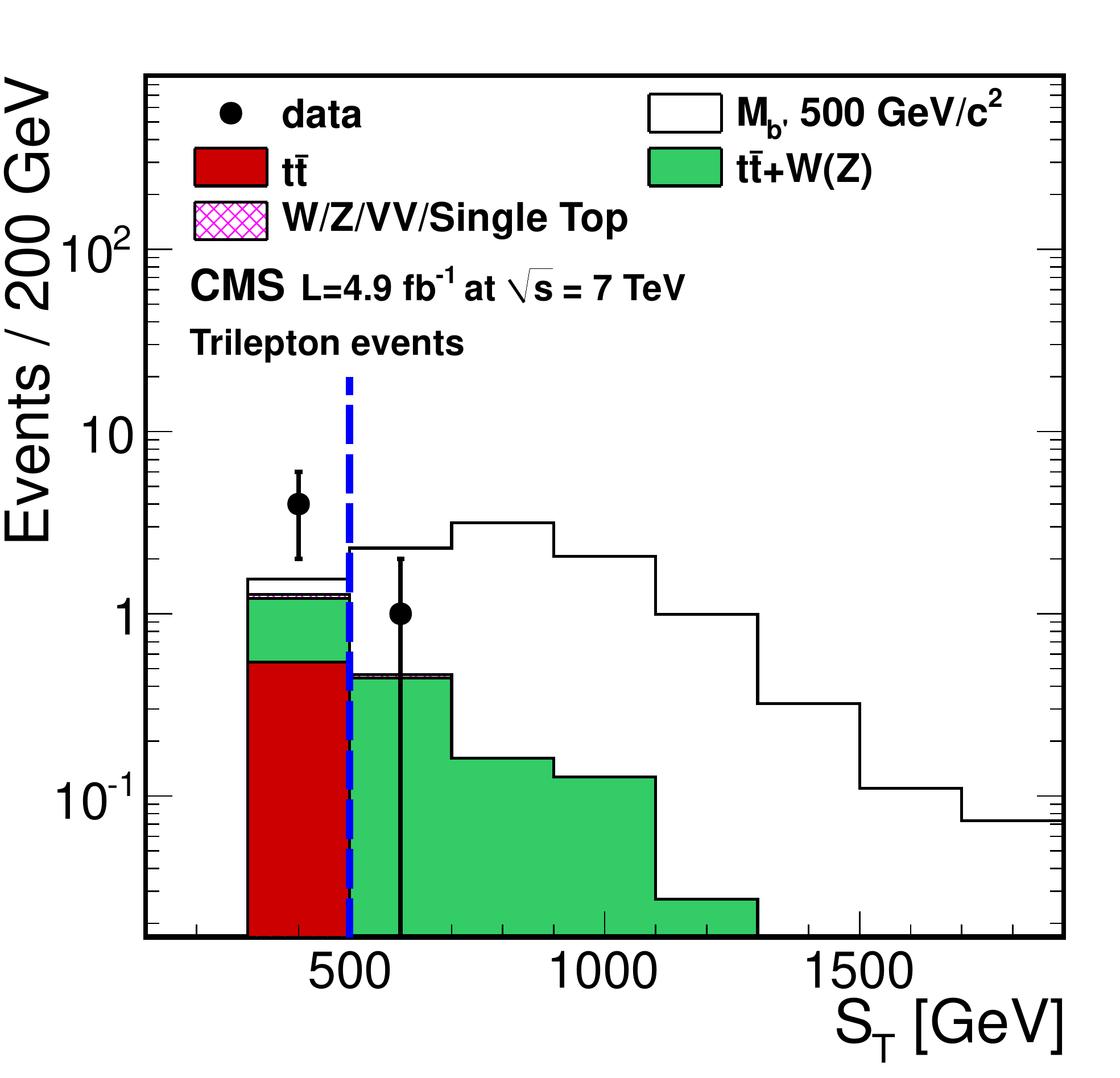}
    \caption{
      Distributions in $S_\mathrm{T}$, the scalar sum of the transverse momenta of objects, in
        the same-charge dilepton channel (left), and the trilepton channel (right).
        The open histogram is the contribution expected from a \cPqbp having $M_{\cPqbp} =500\GeVcc$.
        The histograms for standard model processes are normalized to the total expected background.
        All selection criteria are applied except the one corresponding to the plotted variable.
        The vertical dotted line indicates the lower bound on $S_\mathrm{T}$ used in the analysis.
    }
\label{fig:misc_plots}
\end{center}
\end{figure}

Table~\ref{tab:yield_expected_sig} shows the expected efficiencies
for a \cPqbp signal, for $450 \le M_{\cPqbp} \le 650$\GeVcc.
The efficiencies vary between 1.5\% and 1.7\% for the same-charge dilepton channel,
and between 0.47\% and 0.63\% for the trilepton events,
in the chosen range of $M_{\cPqbp}$.
These efficiencies include the branching fractions for W-decay
and the b-tagging performance~\cite{btagging-performance}.
Jet multiplicities for the same-charge dilepton and the trilepton channels
are shown in Fig.~\ref{fig:jet_multiplicities}, and the $S_\mathrm{T}$ distributions are
presented in Fig.~\ref{fig:misc_plots}.
The expected distributions for a \cPqbp signal having $M_{\cPqbp}=500$\GeVcc
are normalized to the production cross sections from Ref.~\cite{Aliev:2010zk}
that include approximate next-to-next-to-leading-order perturbative QCD corrections, and
standard QCD couplings are assumed.

  \begin{table}[t]
    \caption{Summary of expected $\bpbpbar$ cross sections~\cite{Aliev:2010zk},
    selection efficiencies, and yields for the two signal channels as a function of the \cPqbp mass.
    }
    \label{tab:yield_expected_sig}
    \begin{center}
      \begin{tabular}{|c|c|cc|cc|} \hline
      $M_{\mathrm{\cPqbp}}$ & Cross section & \multicolumn{2}{c|}{Same-charge dilepton} & \multicolumn{2}{c|}{Trilepton} \\
        $[\GeVcc]$ & $[$pb$]$ & efficiency $[\%]$ & yield & efficiency $[\%]$ & yield \\
        \hline
 		450 & 0.662   & $1.52\pm0.13$ & 49  & $0.47\pm0.05$ & 15  \\
 		500 & 0.330   & $1.64\pm0.14$ & 26  & $0.51\pm0.05$ & 8.2 \\
 		550 & 0.171   & $1.71\pm0.14$ & 14  & $0.56\pm0.05$ & 4.7 \\
 		600 & 0.0923  & $1.69\pm0.14$ & 7.6 & $0.60\pm0.06$ & 2.7 \\
 		650 & 0.0511  & $1.71\pm0.15$ & 4.3 & $0.63\pm0.06$ & 1.6 \\

        \hline
      \end{tabular}
    \end{center}
  \end{table}

\section{Background estimation}

Because of the b-tagging requirement,
98\% of the expected background events in the same-charge dilepton channel have at least one top quark from
$\ttbar$, $\ttbar+\PW/\cPZ$, or single-top processes.
These backgrounds are categorized into three sources:
(i) true $\ell^+\ell^-$ events with a electron of misidentified charge,
(ii) single-lepton events with an extra misidentified or non-isolated lepton candidate,
and (iii) events with two prompt leptons of the same charge.
The contribution due to the charge misidentification of electrons is determined using a control sample that, while keeping the remaining signal selection criteria, has oppositely-charged electron pairs or electrons and muons.
The charge misidentification rate (0.03\% and 0.31\% for barrel and endcap candidates, respectively) is
determined by counting the events containing two same-charge electron candidates, whose invariant mass is consistent
with that of a $\cPZ$-boson, relative to the yield of $\cPZ \to\Pep\Pem$ events.
Background from source (ii) is estimated as follows.
Leptons passing the selection criteria described in Section~3 for signal are denoted as ``tight'',
while muon candidates passing relaxed isolation thresholds and track-fit quality requirements,
or electron candidates passing relaxed identification and isolation requirements, are referred to as ``loose''.
Tight lepton candidates are excluded from the selection of loose lepton candidates.
The background from events containing a false or non-isolated
lepton candidate is estimated using another data control sample containing
one tight lepton candidate and one loose lepton candidate,
with the remaining selection criteria kept identical to those used for the signal sample.
By definition, this control sample excludes events in the signal sample.
The contributions of the backgrounds in the selected events
are calculated using the yields observed in the control sample multiplied by the ratio of the number of lepton candidates passing tight selection criteria to those passing the loose criteria. This ratio, also determined in data, is calculated as the number of events containing one loose and one tight lepton candidate divided by the number of those containing two loose lepton candidates.
Applying the above methods to data, a background yield of $7.8\pm2.8$ events is
estimated to originate from sources (i) and (ii).

The estimated yield to the same-charge dilepton channel from processes that produce prompt same-charge dileptons,
including $\ttbar+\cPZ$, $\ttbar+\PW$, and diboson channels
($\PW\cPZ$, $\cPZ\cPZ$, and same-charge $\PW^\pm \PW^\pm+$jets),
is determined using simulations of these processes.
The contribution in the signal region is estimated to be $3.6\pm0.6$ events.

For the trilepton channel,
the background is an order of magnitude smaller than for the same-charge dilepton channel,
and is dominated by processes that produce three prompt leptons, such as $\ttbar+\PW/\cPZ$.
The yield in the signal region, which is only $0.78\pm0.21$ events, is estimated using simulated samples.
Contributions from $\Pp\Pp \to \ttbar$ and $\PW/\cPZ$ processes are normalized to the
cross sections measured by CMS~\cite{Chatrchyan:2011yy, Khachatryan:2010xn}.
The single-top contributions are normalized to the next-to-next-to-leading-logarithm
cross sections~\cite{Kidonakis:2011wy,Kidonakis:2010ux}.
Production rates for dibosons are
estimated from the next-to-leading-order cross sections given by \textsc{mcfm}~\cite{Campbell:2010ff}.
The $\ttbar+\PW/\cPZ$ and
same-charge $\PW^\pm \PW^\pm+$jets processes are normalized
to the next-to-leading-order cross sections given in Ref.~\cite{Hirschi:2011pa}.

The multijet background contribution is estimated using a control sample of events containing
two (three) loose lepton candidates for the same-charge dilepton (trilepton) channel,
maintaining other selection criteria.
The yield of multijet events in the signal region is calculated by multiplying the yield observed
in the control sample by the ratio squared (cubed) of the number of lepton candidates passing tight selection to the number passing loose selection.
The contribution of multijet events to the signal region is estimated to be smaller than 0.12 (0.001) events
for the same-charge dilepton (trilepton) channel, and thus is negligible compared to contributions
from the other background processes.

\section{Systematic uncertainties}

To validate the procedure for estimating background,
and to assign a proper systematic uncertainty,
the study in the same-charge dilepton channel is repeated
using a mixture of simulated samples representing the potential background sources.
The full estimation procedure is then applied to the simulated samples,
and results are compared to the input values.
The observed difference ($2.7\pm0.9$ events) is included as a systematic uncertainty.
The statistical uncertainties
in the control samples are also included in the systematic uncertainties.

The following uncertainties are included in both dilepton and trilepton channels.
The b-tagging efficiency as measured in data has a precision of 10\% per b-jet~\cite{btagging-performance},
resulting in a 6.7\% uncertainty in the efficiency of signal samples.
The effect of this uncertainty on the background contributions determined
using simulated samples is estimated to be 0.35 (0.08) events for the dilepton (trilepton) channel.
Lepton selection efficiencies are measured using
inclusive $\cPZ \to \ell^+\ell^-$ data,
and the difference between efficiencies measured in data and simulation
is taken as a systematic uncertainty.
An additional systematic uncertainty of 50\% of the difference in efficiency
between simulated $\cPZ$ and \cPqbp samples is included,
to cover the effects of different event topologies.
This estimation yields uncertainties of 1.7\% and 2.7\% for electrons and muons, respectively.
The uncertainty in signal efficiency, calculated using appropriate weighting of the electron and muon contributions, is 3.3\% (5.0\%) for the dilepton (trilepton) channel.

The uncertainties in the background normalization are estimated to be
0.74 and 0.12 events for dilepton and trilepton channels, respectively, and the
uncertainties for each of the individual processes are included as follows:
$\pm11$\% for $\ttbar$~\cite{Chatrchyan:2011yy},
$\pm3$\% ($\pm4$\%) for $\PW$ ($\cPZ$)~\cite{Khachatryan:2010xn},
$\pm30$\% for single top processes,
$\pm26$\% for $\PW\PW$, $\pm30$\% for $\PW\cPZ$, $\pm21$\% for $\cPZ\cPZ$,
$\pm30$\% for $\cPqt\cPqt\PW$, $\pm30$\% for $\cPqt\cPqt\cPZ$, $\pm49$\% for $\PW^\pm\PW^\pm+$jets, and $\pm100$\% for multijet.
The uncertainties in the normalization of diboson, $\cPqt\cPqt\PW$, $\cPqt\cPqt\cPZ$, and $\PW^\pm\PW^\pm+$jets processes
are taken from a comparison of next-to-leading-order and leading-order predictions.
The uncertainty related to the presence of additional interactions (pile-up)
in the same beam crossing interval as an event is examined by varying the number of such interactions included in the simulations.
The systematic effects of the uncertainties in jet-energy-scale,
jet resolution, \ETslash resolution,
pile-up events, and trigger efficiency are found to be small~\cite{JES, MET}.
Uncertainty sets given by CTEQ6~\cite{Pumplin:2002vw} are used to determine the
uncertainties from the choice of parton distribution functions.
The relative uncertainty in the
integrated luminosity measurement is estimated to be 2.2\%~\cite{lumi2011},
and is included in the calculation of limits.
The details of uncertainties in the signal selection efficiency and in the
background estimation are presented in Table~\ref{tab:systematic_error}.

  \begin{table}[t]
    \caption{
      Summary of relative systematic uncertainties in signal selection efficiencies ($\Delta\epsilon/\epsilon$) and
        the absolute systematic uncertainties in the number of expected background events ($\Delta B$).
        The ranges given below represent the dependence on $\rm M_{\cPqbp}$, varying from 450\GeVcc to 650\GeVcc.
}
        \label{tab:systematic_error}
    \begin{center}
      \begin{tabular}{|l|cc|cc|} \hline
 & \multicolumn{2}{c|}{Same-charge dilepton} & \multicolumn{2}{c|}{Trilepton} \\
& $\Delta\epsilon/\epsilon$ [\%] & $\Delta B$ & $\Delta\epsilon/\epsilon$ [\%] & $\Delta B$  \\	
	\hline
	Accuracy of control-sample method 	& -          	& 2.63 & -             & -    \\
	Control sample statistics    		& -            	& 0.76 & -            	& -    \\
	\hline	
 	b-tagging             				& 6.7 		    & 0.35 & 6.7 			& 0.08 \\		
	Lepton selection             		& 3.3 	        & 0.03 & 4.9 -- 5.1 	& 0.04 \\
 	Background normalization     		& -            	& 0.74 & -            	& 0.12 \\		
 	Pile-up events                	    & 0.5    	    & 0.13 & 0.6 			& 0.03 \\
 	Jet energy scale             		& 1.1 -- 2.0 	& 0.31 & 0.3 -- 1.0 	& 0.02 \\
 	Jet energy resolution        		& 0.3 -- 1.4 	& 0.22 & 0.3 -- 0.9 	& 0.02 \\
 	Missing energy resolution    		& 0.1 -- 0.7 	& 0.38 & 0.1 -- 2.2 	& 0.07 \\
 	Trigger                      		& 1.4        	& 0.11 & 0.7        	& 0.01 \\
 	PDF                          		& 0.4 -- 1.9 	& 0.26 & 0.4 -- 1.3 	& 0.03 \\
 	Simulated sample statistics  		& 2.7 -- 3.4 	& 0.26 & 4.5 -- 6.5 	& 0.12 \\	
 	Integrated luminosity     			& 2.2        	& 0.24 & 2.2         	& 0.05 \\
	\hline
	Total                        		& 8.6 -- 9.0  	& 2.9  & 10 -- 11   	& 0.21 \\
	\hline
      \end{tabular}
    \end{center}
  \end{table}

\section{Results}

There are 12 (1) events found in the signal region for the dilepton (trilepton) channel,
to be compared with an estimated background of $11.4\pm2.9$ ($0.78\pm0.21$) (Table~\ref{tab:yield_expected_data}).

   \begin{table}[t]
    \caption{Summary of the estimated background contributions to the same-charge dilepton channel and
    the trilepton channel, and the observed event yield in data.
   The given uncertainties are systematic.}
    \label{tab:yield_expected_data}
    \begin{center}
      \begin{tabular}{|l|c|c|} \hline
        \multirow{2}{*}{Sources} & Same-charge & \multirow{2}{*}{Trilepton} \\
                    & dilepton &  \\
        \hline
        Same-charge dilepton with a charge-misidentified electron, or 		& \multirow{2}{*}{$7.8\pm2.8$} & \\
        a misidentified or non-isolated lepton (from data)          	  	&                      & \\
        Prompt same-charge dilepton, or trilepton (simulated)            	& $3.6\pm0.6$          & $0.78\pm0.21$  \\
        \hline
        Background sum              & $11.4\pm2.9$ & $0.78\pm0.21$ \\
        \hline
        Observed yield in data      & 12 & 1 \\
        \hline
      \end{tabular}
    \end{center}
  \end{table}

Most of the background sources contain at least one top quark
in the final state, with a b-quark produced in the top quark decay.
Therefore, modifying the required number of  b-tagged jets, in a separate study, provides a good check of the analysis.
The observed yields when requiring
$\ge 0$, $\ge 1$, or $\ge 2$ b-tagged jets are consistent with the estimated background,
and in agreement with the expected dominance of background from top quarks.

\begin{figure}[tp]
  \begin{center}
    \resizebox{12cm}{!}{\includegraphics{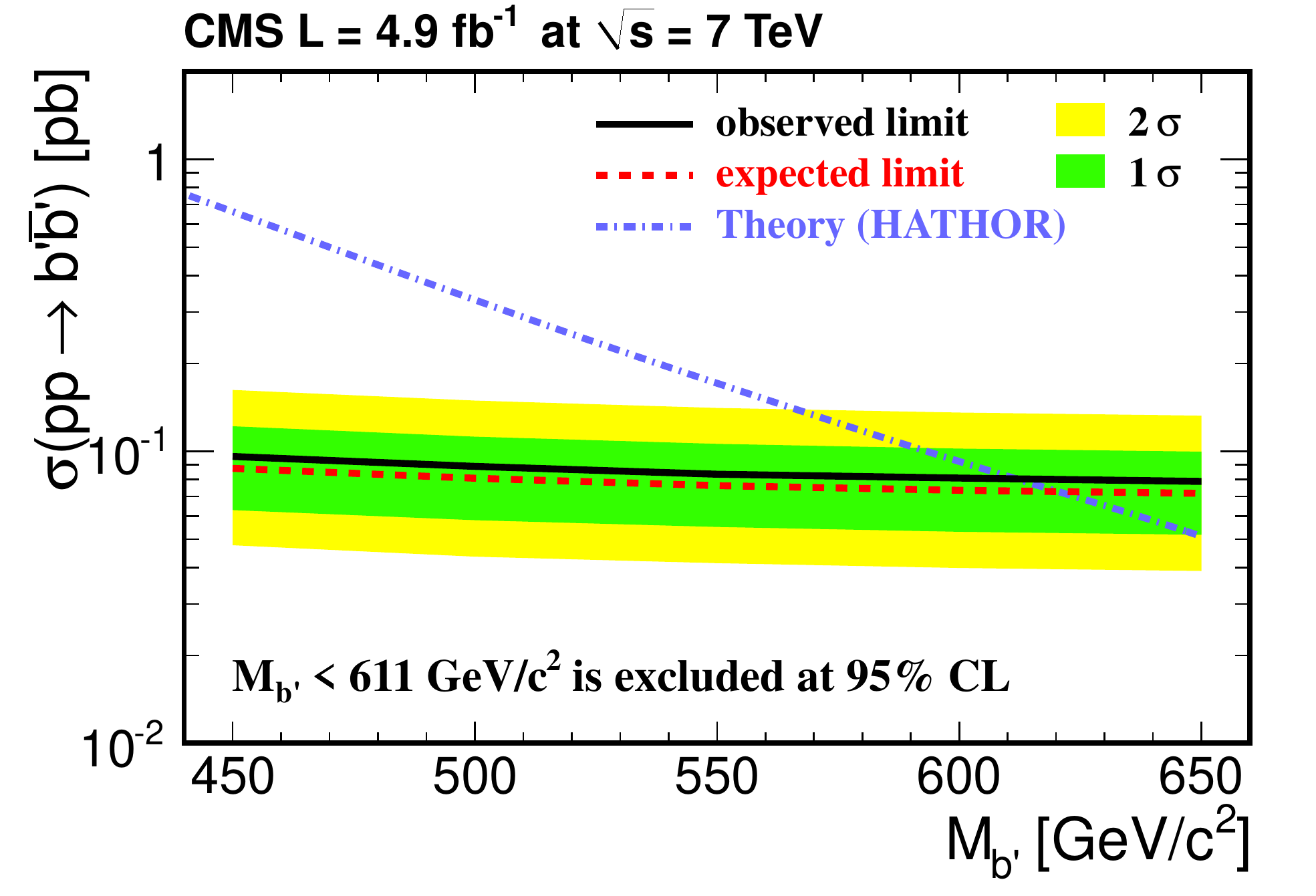}}
    \caption{Exclusion limits at 95\% CL on the $\Pp\Pp \to \bpbpbar$
    production cross section ($\sigma$).
    The solid line represents the observed limits,
    while the dotted line represents
    the limits expected for the available integrated luminosity,
    assuming the presence of standard model processes alone.
    A comparison with the production cross-sections excludes \cPqbp masses
    $M_{\cPqbp} < 611$\GeVcc at 95\% CL
    for a 100\% $\cPqbp \to \cPqt\PW$ decay branching fraction.
    }
    \label{fig:s95exclusion}
  \end{center}
\end{figure}

For each \cPqbp mass hypothesis, cross sections, selection efficiencies, and associated uncertainties are estimated
(Tables~\ref{tab:yield_expected_sig}~and~\ref{tab:systematic_error}).
From these values, the estimated background yield, and the number of observed events,
upper limits on $\bpbpbar$ pair production cross sections at 95\% CL are derived,
using a modified frequentist approach ($CL_s$)~\cite{cls}.
These limits are plotted as the solid line in Fig.~\ref{fig:s95exclusion},
while the dotted line represents the limits expected with the available integrated luminosity,
assuming the presence of standard model processes alone.
By comparing to the theoretical production cross section
for $\Pp\Pp \to \bpbpbar$,
a lower limit of 611\GeVcc is extracted for the mass of the \cPqbp quark, at 95\% CL, while
a limit of 619\GeVcc is expected for a background-only hypothesis.

\section{Summary}

Results have been presented from a search for heavy bottom-like quarks
pair-produced in proton-proton collisions at $\sqrt{s} = 7$\TeV.
The process of $\Pp\Pp \to \bpbpbar \to \ttbar\PWp\PWm$
has been studied in data corresponding to an integrated luminosity of 4.9\fbinv,
collected with the CMS detector.
Estimated background contributions have been found to be small, since final states
containing the signatures of trileptons or same-charge dileptons
are produced rarely in standard model processes.
Assuming a branching fraction of 100\% for the decay $\cPqbp \to \cPqt\PW$, \cPqbp quarks with masses
below 611\GeVcc are excluded at 95\% CL. This is the most stringent limit to date.

\section*{Acknowledgment}

We wish to congratulate our colleagues in the CERN accelerator departments for the excellent performance of the LHC machine. We thank the technical and administrative staff at CERN and other CMS institutes, and acknowledge support from: FMSR (Austria); FNRS and FWO (Belgium); CNPq, CAPES, FAPERJ, and FAPESP (Brazil); MES (Bulgaria); CERN; CAS, MoST, and NSFC (China); COLCIENCIAS (Colombia); MSES (Croatia); RPF (Cyprus); Academy of Sciences and NICPB (Estonia); Academy of Finland, ME, and HIP (Finland); CEA and CNRS/IN2P3 (France); BMBF, DFG, and HGF (Germany); GSRT (Greece); OTKA and NKTH (Hungary); DAE and DST (India); IPM (Iran); SFI (Ireland); INFN (Italy); NRF and WCU (Korea); LAS (Lithuania); CINVESTAV, CONACYT, SEP, and UASLP-FAI (Mexico); PAEC (Pakistan); SCSR (Poland); FCT (Portugal); JINR (Armenia, Belarus, Georgia, Ukraine, Uzbekistan); MST and MAE (Russia); MSTD (Serbia); MICINN and CPAN (Spain); Swiss Funding Agencies (Switzerland); NSC (Taipei); TUBITAK and TAEK (Turkey); STFC (United Kingdom); DOE and NSF (USA).

\bibliography{auto_generated}   

\providecommand{\href}[2]{#2}\begingroup\raggedright\begin{thebibliography}{10}%
\makeatletter
\providecommand{\hrefCMSnoop }[0]{\@secondoftwo}%
\makeatother
\providecommand{\doi}{\texttt{doi:}\begingroup \urlstyle{tt}\Url}

\bibitem{Holdom:2009rf}
B.~Holdom\hrefCMSnoop {} { {et~al.}, ``Four Statements about the Fourth
  Generation'',} \textit{ PMC Phys. A} \textbf{ 3} (2009) 4,
  \href{http://dx.doi.org/10.1186/1754-0410-3-4}{\doi{10.1186/1754-0410-3-4}},
\href{http://www.arXiv.org/abs/0904.4698}{\texttt{ arXiv:0904.4698}}.

\bibitem{Soni:2010xh}
A.~Soni\hrefCMSnoop {} { {et~al.}, ``{SM} with four generations: Selected
  implications for rare {B} and {K} decays'',} \textit{ Phys. Rev. D} \textbf{
  82} (2010) 033009,
  \href{http://dx.doi.org/10.1103/PhysRevD.82.033009}{\doi{10.1103/PhysRevD.82.033009}},
\href{http://www.arXiv.org/abs/1002.0595}{\texttt{ arXiv:1002.0595}}.

\bibitem{Buras:2010pi}
A.~J. Buras\hrefCMSnoop {} { {et~al.}, ``Patterns of flavour violation in the
  presence of a fourth generation of quarks and leptons'',} \textit{ JHEP}
  \textbf{ 09} (2010) 106,
  \href{http://dx.doi.org/10.1007/JHEP09(2010)106}{\doi{10.1007/JHEP09(2010)106}},
\href{http://www.arXiv.org/abs/1002.2126}{\texttt{ arXiv:1002.2126}}.

\bibitem{Erler:2010sk}
\hrefCMSnoop {} {J.~Erler and P.~Langacker, ``Precision Constraints on Extra
  Fermion Generations'',} \textit{ Phys. Rev. Lett.} \textbf{ 105} (2010)
  031801,
  \href{http://dx.doi.org/10.1103/PhysRevLett.105.031801}{\doi{10.1103/PhysRevLett.105.031801}},
\href{http://www.arXiv.org/abs/1003.3211}{\texttt{ arXiv:1003.3211}}.

\bibitem{Flacco:2010rg}
C.~J. Flacco\hrefCMSnoop {} { {et~al.}, ``Direct Mass Limits for Chiral
  Fourth-Generation Quarks in All Mixing Scenarios'',} \textit{ Phys. Rev.
  Lett.} \textbf{ 105} (2010) 111801,
  \href{http://dx.doi.org/10.1103/PhysRevLett.105.111801}{\doi{10.1103/PhysRevLett.105.111801}},
\href{http://www.arXiv.org/abs/1005.1077}{\texttt{ arXiv:1005.1077}}.

\bibitem{Denner:2011vt}
A.~Denner\hrefCMSnoop {} { {et~al.}, ``{Higgs production and decay with a
  fourth Standard-Model-like fermion generation}'',} (2011).
\href{http://www.arXiv.org/abs/1111.6395}{\texttt{ arXiv:1111.6395}}.

\bibitem{Frampton:1999xi}
\hrefCMSnoop {} {P.~H. Frampton, P.~Q. Hung, and M.~Sher, ``Quarks and leptons
  beyond the third generation'',} \textit{ Phys. Rept.} \textbf{ 330} (2000)
  263,
  \href{http://dx.doi.org/10.1016/S0370-1573(99)00095-2}{\doi{10.1016/S0370-1573(99)00095-2}},
\href{http://www.arXiv.org/abs/hep-ph/9903387}{\texttt{ arXiv:hep-ph/9903387}}.

\bibitem{Kribs:2007nz}
G.~D. Kribs\hrefCMSnoop {} { {et~al.}, ``Four generations and {Higgs}
  physics'',} \textit{ Phys. Rev. D} \textbf{ 76} (2007) 075016,
  \href{http://dx.doi.org/10.1103/PhysRevD.76.075016}{\doi{10.1103/PhysRevD.76.075016}},
\href{http://www.arXiv.org/abs/0706.3718}{\texttt{ arXiv:0706.3718}}.

\bibitem{Hou:2008xd}
\hrefCMSnoop {} {W.-S. Hou, ``Source of {CP} Violation for the Baryon Asymmetry
  of the Universe'',} \textit{ Chin. J. Phys.} \textbf{ 47} (2009) 134,
\href{http://www.arXiv.org/abs/0803.1234}{\texttt{ arXiv:0803.1234}}.

\bibitem{Decamp:1989tu}
\hrefCMSnoop {} {{ ALEPH} Collaboration, ``Determination of the number of light
  neutrino species'',} \textit{ Phys. Lett. B} \textbf{ 231} (1989) 519,
\href{http://dx.doi.org/10.1016/0370-2693(89)90704-1}{\doi{10.1016/0370-2693(89)90704-1}}.

\bibitem{Aarnio:1989tv}
\hrefCMSnoop {} {{ DELPHI} Collaboration, ``Measurement of the mass and width
  of the $\rm {Z}^0$ particle from multi-hadronic final states produced in $\rm
  e^{+} e^{-}$ annihilations'',} \textit{ Phys. Lett. B} \textbf{ 231} (1989)
  539,
\href{http://dx.doi.org/10.1016/0370-2693(89)90706-5}{\doi{10.1016/0370-2693(89)90706-5}}.

\bibitem{Adeva:1989mn}
\hrefCMSnoop {} {{ L3} Collaboration, ``A determination of the properties of
  the neutral intermediate vector boson $\rm {Z}^0$'',} \textit{ Phys. Lett. B}
  \textbf{ 231} (1989) 509,
\href{http://dx.doi.org/10.1016/0370-2693(89)90703-X}{\doi{10.1016/0370-2693(89)90703-X}}.

\bibitem{Akrawy:1989pi}
\hrefCMSnoop {} {{ OPAL} Collaboration, ``Measurement of the $\rm {Z}^0$ mass
  and width with the {OPAL} detector at {LEP}'',} \textit{ Phys. Lett. B}
  \textbf{ 231} (1989) 530,
\href{http://dx.doi.org/10.1016/0370-2693(89)90705-3}{\doi{10.1016/0370-2693(89)90705-3}}.

\bibitem{Aad:2012us}
\hrefCMSnoop {} {{ ATLAS} Collaboration, ``{Search for down-type fourth
  generation quarks with the ATLAS detector in events with one lepton and high
  transverse momentum hadronically decaying W bosons in sqrt(s) = 7 TeV pp
  collisions}'',} (2012). \href{http://www.arXiv.org/abs/1202.6540}{\texttt{
  arXiv:1202.6540}}.
Submitted to \textit{Phys. Rev. Lett.}

\bibitem{Aaltonen:2009nr}
\hrefCMSnoop {} {{ CDF} Collaboration, ``Search for New Bottomlike Quark Pair
  Decays $Q\overline{Q} \to (tW^{\mp})(\overline{t}W^{\pm})$ in Same-Charge
  Dilepton Events'',} \textit{ Phys. Rev. Lett.} \textbf{ 104} (2010) 091801,
  \href{http://dx.doi.org/10.1103/PhysRevLett.104.091801}{\doi{10.1103/PhysRevLett.104.091801}},
\href{http://www.arXiv.org/abs/0912.1057}{\texttt{ arXiv:0912.1057}}.

\bibitem{Aaltonen:2011vr}
\hrefCMSnoop {} {{ CDF} Collaboration, ``Search for heavy bottom-like quarks
  decaying to an electron or muon and jets in $p\bar{p}$ collisions at
  $\sqrt{s}=1.96$ TeV'',} \textit{ Phys. Rev. Lett.} \textbf{ 106} (2011)
  141803,
  \href{http://dx.doi.org/10.1103/PhysRevLett.106.141803}{\doi{10.1103/PhysRevLett.106.141803}},
\href{http://www.arXiv.org/abs/1101.5728}{\texttt{ arXiv:1101.5728}}.

\bibitem{Chatrchyan:2011em}
\hrefCMSnoop {} {{ CMS} Collaboration, ``{Search for a heavy bottom-like quark
  in pp collisions at $\sqrt{s} = 7$ TeV}'',} \textit{ Phys. Lett. B} \textbf{
  B701} (2011) 204,
  \href{http://dx.doi.org/10.1016/j.physletb.2011.05.074}{\doi{10.1016/j.physletb.2011.05.074}},
\href{http://www.arXiv.org/abs/1102.4746}{\texttt{ arXiv:1102.4746}}.

\bibitem{Abazov:2011vy}
\hrefCMSnoop {} {{ D0} Collaboration, ``{Search for a fourth generation $t'$
  quark in $p\bar{p}$ collisions at $\sqrt{s}=1.96$ TeV}'',} (2011).
  \href{http://www.arXiv.org/abs/1104.4522}{\texttt{ arXiv:1104.4522}}.
Submitted to \textit{Phys. Rev. Lett.}

\bibitem{Aad:2012bt}
\hrefCMSnoop {} {{ ATLAS} Collaboration, ``{Search for pair-produced heavy
  quarks decaying to Wq in the two-lepton channel at sqrt(s) = 7 TeV with the
  ATLAS detector}'',} \href{http://www.arXiv.org/abs/1202.3389}{\texttt{
  arXiv:1202.3389}}.
Submitted to \textit{Phys. Rev. D}.

\bibitem{Aad:2012xc}
\hrefCMSnoop {} {{ ATLAS} Collaboration, ``{Search for pair production of a
  heavy quark decaying to a W boson and a b quark in the lepton+jets channel
  with the ATLAS detector}'',}
  \href{http://www.arXiv.org/abs/1202.3076}{\texttt{ arXiv:1202.3076}}.
Submitted to \textit{Phys. Rev. Lett.}

\bibitem{CMS_tprime_dilepton}
\hrefCMSnoop {} {{ CMS} Collaboration, ``{Search for heavy, top-like quark pair
  production in the dilepton final state in pp collisions at sqrt(s) = 7
  TeV}'',} \href{http://www.arXiv.org/abs/1203.5410}{\texttt{
  arXiv:1203.5410}}.
Submitted to \textit{Phys. Lett. B}.

\bibitem{Arhrib:2006pm}
\hrefCMSnoop {} {A.~Arhrib and W.-S. Hou, ``Flavor changing neutral currents
  involving heavy quarks with four generations'',} \textit{ JHEP} \textbf{ 07}
  (2006) 009,
  \href{http://dx.doi.org/10.1088/1126-6708/2006/07/009}{\doi{10.1088/1126-6708/2006/07/009}},
\href{http://www.arXiv.org/abs/hep-ph/0602035}{\texttt{ arXiv:hep-ph/0602035}}.

\bibitem{:2008zzk}
\hrefCMSnoop {} {{ CMS} Collaboration, ``The {CMS} experiment at the {CERN
  LHC}'',} \textit{ JINST} \textbf{ 03} (2008) S08004,
\href{http://dx.doi.org/10.1088/1748-0221/3/08/S08004}{\doi{10.1088/1748-0221/3/08/S08004}}.

\bibitem{Adam:2005zf}
\hrefCMSnoop {} {{ CMS} Collaboration, ``The {CMS} high level trigger'',}
  \textit{ Eur. Phys. J. C} \textbf{ 46} (2006) 605,
  \href{http://dx.doi.org/10.1140/epjc/s2006-02495-8}{\doi{10.1140/epjc/s2006-02495-8}},
\href{http://www.arXiv.org/abs/hep-ex/0512077}{\texttt{ arXiv:hep-ex/0512077}}.

\bibitem{PFT-09-001}
\href {http://cdsweb.cern.ch/record/1194487} {{ CMS} Collaboration,
  ``Particle-flow event reconstruction in {CMS} and performance for jets, taus,
  and {MET}'',} CMS Physics Analysis Summary CMS-PAS-PFT-09-001, (2009).

\bibitem{PFT-10-001}
\href {http://cdsweb.cern.ch/record/1247373} {{ CMS} Collaboration,
  ``Commissioning of the Particle-flow Event Reconstruction with the first
  {LHC} collisions recorded in the {CMS} detector'',} CMS Physics Analysis
  Summary CMS-PAS-PFT-10-001, (2010).

\bibitem{PFJETPAS}
\href {http://cdsweb.cern.ch/record/1279341} {{ CMS} Collaboration,
  ``Commissioning of the Particle-Flow Reconstruction in Minimum-Bias and Jet
  Events from {\Pp\Pp} Collisions at 7 {TeV}'',} CMS Physics Analysis Summary
  CMS-PAS-PFT-10-002, (2010).

\bibitem{PFT-10-003}
\href {http://cdsweb.cern.ch/record/1279347} {{ CMS} Collaboration,
  ``Commissioning of the particle-flow event reconstruction with leptons from
  {J}/$\Psi$ and {W} decays at 7 {TeV}'',} CMS Physics Analysis Summary
  CMS-PAS-PFT-10-003, (2010).

\bibitem{MUOPAS}
\href {http://cdsweb.cern.ch/record/1279140} {{ CMS} Collaboration,
  ``Performance of muon identification in pp collisions at $\sqrt{s}$ = 7
  {TeV}'',} CMS Physics Analysis Summary CMS-PAS-MUO-10-002, (2010).

\bibitem{EGMPAS}
\href {http://cdsweb.cern.ch/record/1299116} {{ CMS} Collaboration, ``Electron
  Reconstruction and Identification at $\sqrt{s} = 7$ {TeV}'',} CMS Physics
  Analysis Summary CMS-PAS-EGM-10-004, (2010).

\bibitem{Cacciari:2008gp}
\hrefCMSnoop {} {M.~Cacciari, G.~P. Salam, and G.~Soyez, ``The anti-$k_t$ jet
  clustering algorithm'',} \textit{ JHEP} \textbf{ 04} (2008) 063,
  \href{http://dx.doi.org/10.1088/1126-6708/2008/04/063}{\doi{10.1088/1126-6708/2008/04/063}},
\href{http://www.arXiv.org/abs/0802.1189}{\texttt{ arXiv:0802.1189}}.

\bibitem{JES}
\hrefCMSnoop {} {{CMS collaboration}, ``{Determination of jet energy
  calibration and transverse momentum resolution in CMS}'',} \textit{ JINST}
  \textbf{ 6} (2011) 11002,
  \href{http://dx.doi.org/10.1088/1748-0221/6/11/P11002}{\doi{10.1088/1748-0221/6/11/P11002}},
  \href{http://www.arXiv.org/abs/1107.4277}{\texttt{ arXiv:1107.4277}}.

\bibitem{MET}
\hrefCMSnoop {} {{CMS collaboration}, ``{Missing transverse energy performance
  of the CMS detector}'',} \textit{ JINST} \textbf{ 6} (2011) 9001,
  \href{http://dx.doi.org/10.1088/1748-0221/6/09/P09001}{\doi{10.1088/1748-0221/6/09/P09001}},
  \href{http://www.arXiv.org/abs/1106.5048}{\texttt{ arXiv:1106.5048}}.

\bibitem{btagging-performance}
\href {http://cdsweb.cern.ch/record/1366061} {{ CMS} Collaboration,
  ``Performance of b-jet identification in {CMS}'',} CMS Physics Analysis
  Summary CMS-PAS-BTV-11-001, (2011).

\bibitem{2010EPJC..tmp..299K}
\hrefCMSnoop {} {{ CMS} Collaboration, ``{CMS} tracking performance results
  from early {LHC} operation'',} \textit{ Eur. Phys. J.} \textbf{ C70} (2010)
  1165,
  \href{http://dx.doi.org/10.1140/epjc/s10052-010-1491-3}{\doi{10.1140/epjc/s10052-010-1491-3}},
  \href{http://www.arXiv.org/abs/1007.1988}{\texttt{ arXiv:1007.1988}}.

\bibitem{Maltoni:2002qb}
\hrefCMSnoop {} {F.~Maltoni and T.~Stelzer, ``{MadEvent}: Automatic event
  generation with {MadGraph}'',} \textit{ JHEP} \textbf{ 02} (2003) 027,
  \href{http://dx.doi.org/10.1088/1126-6708/2003/02/027}{\doi{10.1088/1126-6708/2003/02/027}},
\href{http://www.arXiv.org/abs/hep-ph/0208156}{\texttt{ arXiv:hep-ph/0208156}}.

\bibitem{PYTHIA}
\hrefCMSnoop {} {T.~Sj{\"o}strand, S.~Mrenna, and P.~Skands, ``{PYTHIA} 6.4
  Physics and Manual'',} \textit{ JHEP} \textbf{ 05} (2006) 026,
  \href{http://dx.doi.org/10.1088/1126-6708/2006/05/026}{\doi{10.1088/1126-6708/2006/05/026}},
  \href{http://www.arXiv.org/abs/hep-ph/0603175}{\texttt{
  arXiv:hep-ph/0603175}}.

\bibitem{Hoche:2006ph}
\hrefCMSnoop {} {S.~Hoeche {et~al.}, ``Matching parton showers and matrix
  elements'',} (2006).
\href{http://www.arXiv.org/abs/hep-ph/0602031}{\texttt{ arXiv:hep-ph/0602031}}.

\bibitem{Agostinelli:2002hh}
\hrefCMSnoop {} {{ GEANT4} Collaboration, ``{GEANT4}--a simulation toolkit'',}
  \textit{ Nucl. Instrum. Meth.} \textbf{ A506} (2003) 250,
\href{http://dx.doi.org/10.1016/S0168-9002(03)01368-8}{\doi{10.1016/S0168-9002(03)01368-8}}.

\bibitem{Aliev:2010zk}
\hrefCMSnoop {} {M.~Aliev {et~al.}, ``{HATHOR -- HAdronic Top and Heavy quarks
  crOss section calculatoR}'',} \textit{ Comput. Phys. Commun.} \textbf{ 182}
  (2011) 1034,
  \href{http://dx.doi.org/10.1016/j.cpc.2010.12.040}{\doi{10.1016/j.cpc.2010.12.040}},
\href{http://www.arXiv.org/abs/1007.1327}{\texttt{ arXiv:1007.1327}}.

\bibitem{Chatrchyan:2011yy}
\hrefCMSnoop {} {{ CMS} Collaboration, ``{Measurement of the $t\bar{t}$
  production cross section in pp collisions at 7 TeV in lepton + jets events
  using b-quark jet identification}'',} \textit{ Phys. Rev. D} \textbf{ 84}
  (2011) 092004,
  \href{http://dx.doi.org/10.1103/PhysRevD.84.092004}{\doi{10.1103/PhysRevD.84.092004}},
\href{http://www.arXiv.org/abs/1108.3773}{\texttt{ arXiv:1108.3773}}.

\bibitem{Khachatryan:2010xn}
\hrefCMSnoop {} {{ CMS} Collaboration, ``Measurements of inclusive {W} and {Z}
  cross sections in pp collisions at $\sqrt{s}=7$ {TeV}'',} \textit{ JHEP}
  \textbf{ 01} (2011) 080,
  \href{http://dx.doi.org/10.1007/JHEP01(2011)080}{\doi{10.1007/JHEP01(2011)080}},
\href{http://www.arXiv.org/abs/1012.2466}{\texttt{ arXiv:1012.2466}}.

\bibitem{Kidonakis:2011wy}
\hrefCMSnoop {} {N.~Kidonakis, ``{Next-to-next-to-leading-order collinear and
  soft gluon corrections for t-channel single top quark production}'',}
  \textit{ Phys. Rev. D} \textbf{ 83} (2011) 091503,
  \href{http://dx.doi.org/10.1103/PhysRevD.83.091503}{\doi{10.1103/PhysRevD.83.091503}},
  \href{http://www.arXiv.org/abs/1103.2792}{\texttt{ arXiv:1103.2792}}.

\bibitem{Kidonakis:2010ux}
\hrefCMSnoop {} {N.~Kidonakis, ``{Two-loop soft anomalous dimensions for single
  top quark associated production with a W- or H-}'',} \textit{ Phys. Rev.}
  \textbf{ D82} (2010) 054018,
  \href{http://dx.doi.org/10.1103/PhysRevD.82.054018}{\doi{10.1103/PhysRevD.82.054018}},
\href{http://www.arXiv.org/abs/1005.4451}{\texttt{ arXiv:1005.4451}}.

\bibitem{Campbell:2010ff}
\hrefCMSnoop {} {J.~M. Campbell and R.~K. Ellis, ``{MCFM} for the {Tevatron}
  and the {LHC}'',} \textit{ Nucl. Phys. Proc. Suppl.} \textbf{ 205-206} (2010)
  10,
  \href{http://dx.doi.org/10.1016/j.nuclphysbps.2010.08.011}{\doi{10.1016/j.nuclphysbps.2010.08.011}},
\href{http://www.arXiv.org/abs/1007.3492}{\texttt{ arXiv:1007.3492}}.

\bibitem{Hirschi:2011pa}
V.~Hirschi\hrefCMSnoop {} { {et~al.}, ``{Automation of one-loop QCD
  corrections}'',} \textit{ JHEP} \textbf{ 05} (2011) 044,
  \href{http://dx.doi.org/10.1007/JHEP05(2011)044}{\doi{10.1007/JHEP05(2011)044}},
\href{http://www.arXiv.org/abs/1103.0621}{\texttt{ arXiv:1103.0621}}.

\bibitem{Pumplin:2002vw}
J.~Pumplin\hrefCMSnoop {} { {et~al.}, ``New generation of parton distributions
  with uncertainties from global QCD analysis'',} \textit{ JHEP} \textbf{ 07}
  (2002) 012,
\href{http://www.arXiv.org/abs/hep-ph/0201195}{\texttt{ arXiv:hep-ph/0201195}}.

\bibitem{lumi2011}
\href {http://cdsweb.cern.ch/record/1376102} {{ CMS} Collaboration, ``Absolute
  Calibration of the {CMS} Luminosity Measurement: {S}ummer 2011 Update'',} CMS
  Physics Analysis Summary CMS-PAS-EWK-11-001, (2011).

\bibitem{cls}
\hrefCMSnoop {} {A.~L. Read, ``{Presentation of search results: the $CL_s$
  technique}'',} \textit{ J. Phys. G} \textbf{ 28} (2002) 2693,
\href{http://dx.doi.org/10.1088/0954-3899/28/10/313}{\doi{10.1088/0954-3899/28/10/313}}.

\end{thebibliography}\endgroup

\cleardoublepage \appendix\section{The CMS Collaboration \label{app:collab}}\begin{sloppypar}\hyphenpenalty=5000\widowpenalty=500\clubpenalty=5000\textbf{Yerevan Physics Institute,  Yerevan,  Armenia}\\*[0pt]
S.~Chatrchyan, V.~Khachatryan, A.M.~Sirunyan, A.~Tumasyan
\vskip\cmsinstskip
\textbf{Institut f\"{u}r Hochenergiephysik der OeAW,  Wien,  Austria}\\*[0pt]
W.~Adam, T.~Bergauer, M.~Dragicevic, J.~Er\"{o}, C.~Fabjan, M.~Friedl, R.~Fr\"{u}hwirth, V.M.~Ghete, J.~Hammer\cmsAuthorMark{1}, N.~H\"{o}rmann, J.~Hrubec, M.~Jeitler, W.~Kiesenhofer, M.~Krammer, D.~Liko, I.~Mikulec, M.~Pernicka$^{\textrm{\dag}}$, B.~Rahbaran, C.~Rohringer, H.~Rohringer, R.~Sch\"{o}fbeck, J.~Strauss, A.~Taurok, F.~Teischinger, P.~Wagner, W.~Waltenberger, G.~Walzel, E.~Widl, C.-E.~Wulz
% ...
\vskip\cmsinstskip
\dag:~Deceased\\
1:~~Also at CERN, European Organization for Nuclear Research, Geneva, Switzerland\\
% ...

\end{sloppypar}
\end{document}